\documentclass[superscriptaddress,aps,amsfonts,notitlepage,twocolumn, nofootinbib]{revtex4-1}
\usepackage{xcolor,graphicx}
\usepackage{breqn,amsmath,amssymb}

\usepackage{enumerate}

\usepackage{footmisc}

\usepackage{booktabs}
\usepackage{makecell}

\usepackage{comment}

\usepackage{subcaption}

\usepackage[T1]{fontenc}
\usepackage{natbib,hyperref}

\usepackage{mathtools}

\usepackage{centernot}

\makeatletter
\let\cat@comma@active\@empty
\makeatother

\usepackage{IEEEtrantools}

\begin{document}

\bstctlcite{IEEEexample:BSTcontrol}

\title{Stability of nonsingular cosmologies in Galileon models with torsion. A no-go theorem for eternal subluminality}

\author{S. Mironov}
\email{sa.mironov\_1@physics.msu.ru}
\affiliation{Institute for Nuclear Research of the Russian Academy of Sciences, 
60th October Anniversary Prospect, 7a, 117312 Moscow, Russia}
\affiliation{Institute for Theoretical and Mathematical Physics,
MSU, 119991 Moscow, Russia}
\affiliation{NRC, "Kurchatov Institute", 123182, Moscow, Russia}

\author{M. Valencia-Villegas}
\email{mvalenciavillegas@itmp.msu.ru}
\affiliation{Institute for Theoretical and Mathematical Physics,
MSU, 119991 Moscow, Russia}

\begin{abstract}
Generic models in Galileons or Horndeski theory do not have cosmological solutions  that are free of instabilities and singularities in the entire time of evolution. We extend this No-Go theorem to a spacetime with torsion. On this more general geometry the No-Go argument now holds provided the additional hypothesis that the graviton is also subluminal throughout the entire evolution. Thus, critically different for Galileons' stability on a torsionful spacetime, an  arguably unphysical although    arbitrarily short (deep UV) phase occurring at an arbitrary time, when the speed of gravity $(c_g)$ is slightly higher than luminal $(c)$, and by at least an amount $c_g\geq \,\sqrt{2}\,c $, can lead to an all-time linearly stable and nonsingular cosmology. As a proof of principle we build a stable model for a cosmological bounce that is almost always subluminal, where the short-lived superluminal phase occurs  before the bounce and that transits to General Relativity in the asymptotic past and future.
\end{abstract}

\maketitle

Galileons is a well motivated modification of General Relativity by a scalar, with higher derivatives in the action but with second order equations of motion \cite{nicolis2009galileon}. The generalization is equivalent to  Horndeski theory \cite{Deffayet:2011gz,horndeski1974second,Kobayashi:2011nu} and it has nonsingular solutions that generally suffer of gradient instabilities at some time in the entire evolution, up to some special cases  \cite{Rubakov:2016zah,Libanov:2016kfc, Kobayashi:2016xpl,Kolevatov:2016ppi,Mironov:2019fop,Akama:2017jsa,Creminelli:2016zwa, Mironov:2022quk, Evslin:2011vh}. Although these pathologies can happen away from the physically relevant phase, a conclusive resolution to this issue at all times in generic models seems unlikely because the No-Go argument for stability also holds with very general extra matter \cite{Kolevatov:2016ppi,Akama:2017jsa,Creminelli:2016zwa}. We extend this No-Go theorem to a spacetime with torsion in section \ref{sec main} and show that a torsionful geometry may  support stable solutions in Galileons if there exists a superluminal phase. It can {\it formally} happen as a  deep UV inconsistency at an arbitrary time, namely, it can be arbitrarily short and unrelated to the much longer physically relevant length scales, {\it e.g.} width of a bounce, but it casts doubt on Lorentz invariant UV-completions \cite{Adams:2006sv,Dubovsky:2005xd, Creminelli:2022onn}.  In section \ref{sec bounce} we build a toy model for a bounce that is {\it always} stable.

\section{EXTENSION OF THE NO-GO THEOREM TO GALILEONS ON A SPACETIME WITH TORSION}\label{sec main}

We consider up to quartic Generalized Galileons in the notation of \cite{Kobayashi:2011nu} and we assume a spacetime with torsion:
\begin{dmath}
\mathcal{S}=\int \,\text{d}^4x\,\sqrt{-g}\,\left(G_2\,-G_{3}\, {\tilde{\nabla}}_\mu{\tilde{\nabla}}^\mu\phi\,+\,G_4\,\tilde{R}+G_{4,X}\left(\left(\tilde{\nabla}_\mu\tilde{\nabla}^\mu\phi\right)^2-\left(\tilde{\nabla}_\mu\tilde{\nabla}_\nu\phi\right) \tilde{\nabla}^\nu\tilde{\nabla}^\mu\phi \right)\right)\,,\label{eqn HC234c0action}
\end{dmath}
where $G_2,\, G_{3},\, G_4$ are arbitrary functions of  $\phi$ and $X=-\frac{1}{2}g^{\mu\nu}\partial_\mu\phi \partial_\nu\phi$, $g$ is the determinant of the metric with mostly $+$ signature,  $G_{4,X}=\partial G_4/\partial X$ and $\tilde{R}$ and $\tilde{\nabla}$ denote the Ricci scalar and covariant derivative computed with a torsionful connection. The $G_3$ term is the simplest one to "feel" the torsion on the spacetime. For $G_4$ let us stress on the specific order of contraction of Lorentz indices in the last term in (\ref{eqn HC234c0action}). Indeed, second covariant derivatives with torsion do not commute on a scalar, and it was found in \cite{Mironov:2023kzt} that this choice is the only  one that leads to a scalar with a wave-like dispersion relation, as in torsionless Galileons. Hence, (\ref{eqn HC234c0action}) is the relevant choice to the question of how a different geometry can help to the stability of the usual Galileon degrees of freedom.

\paragraph{Quadratic action for Galileons on a torsionful vs. torsionless spacetime:}
We analyze the stability of the FLRW background against linearized perturbations. A straightforward computation shows that all independent components of torsion in (\ref{eqn HC234c0action}) are non dynamical (See Appendix \ref{sec notation} for a detailed derivation). The important aspect is that we can cast the quadratic action of Galileons with torsion in a form reminiscent of the usual Galileons without torsion \cite{Kobayashi:2011nu}. Namely, from (\ref{eqn HC234c0action})

\begin{equation}
\resizebox{.9\hsize}{!}{$\mathcal{S}_{\tau} =\int\, \textrm{d}\eta\,\textrm{d}^3x \,a^4\,\left[\frac{1}{2\,a^2}\left({\mathcal{G}_\tau}\left(\dot{h}_{ij}\right)^2-{\mathcal{F}_\tau}(\partial_k\,{h}_{ij})^2\right)\right]$}\,\label{eqn qlf tensor}
\end{equation}
is the action for the graviton $h_{ij}$ with speed $c_g^2= \mathcal{F}_\tau / \mathcal{G}_\tau$, where $\eta$ is conformal time. The vector sector is non dynamical. The scalar sector reads in the unitary gauge
\begin{dmath}
\mathcal{S}_{s}\,\text{$=$}\, \int\, \textrm{d}\eta\,\textrm{d}^3x\, a^4 \left(- 3\,\frac{ \, \mathcal{G}_{\tau}}{a^2}\,\dot{\psi}^2 \,+ \frac{ \,\mathcal{F}_{\tau}}{a^2}\, \,({{\partial_i \psi}})^2 +6\,\frac{ \Theta}{a}\,\alpha\,\dot{\psi}\,+ 2\, \frac{T}{a^2} \,{{\partial_i \alpha}} \,{{\partial_i \psi}} \,+ 2\frac{{{\partial_i \partial_i B}}}{a^2}\,\left(\,a\, \Theta \,{{\alpha}}\,-\, \, \mathcal{G}_{\tau}\, \,{\dot{\psi}}\right) \,+\,\Sigma\,\alpha^2 \right)\,\,,\nonumber\\
\label{eqn ql0 scalar}
\end{dmath}
where $\psi,\, \alpha$ and $B$ are scalar perturbations and
\begin{eqnarray} 
\begin{array}{ccc}
\mathcal{G}_\tau = 2\frac{\, G_4^2}{\,G_4+\,2\,X \, G_{4,X}}\,, \,& \mathcal{F}_\tau =2\,G_4\,, & T= \mathcal{F}_\tau\,(c_g^2-2)\,,\label{eqn T}
\end{array}
\end{eqnarray}
\begin{eqnarray}
\begin{array}{cc}
\Theta= \frac{4\,\mathcal{G}_\tau^2\, \theta}{\,a\, \mathcal{F}_\tau^4}\,,\label{eqn Theta} &
\Sigma= \frac{2\,\mathcal{G}_\tau^3\,\sigma}{ \,{a} \, \mathcal{F}_\tau^6}\,,\label{eqn Sigma}
\end{array}
\end{eqnarray}
$\theta $ and $\sigma$ (shown in the Appendix \ref{sec app Coeff final action}) depend on two background fields: the scale factor of the FLRW metric $a(\eta)$ and the Galileon/ Horndeski scalar, which in the context of linearized expressions we also denote as $\phi(\eta)$ and $X=\frac{\dot{\phi}^2}{2\,a^2}$. Let us also note that there is a non-trivial torsion background $x(\eta)$ expressed by the background equations in terms of $a(\eta),\, \phi(\eta)$. We show these details in the Appendix \ref{sec notation} for completeness.

Let us notice that despite the similarities between the quadratic actions in torsionless and torsionful Galileons, there is a crucial difference in (\ref{eqn ql0 scalar}) that helps to the stability of the theory with torsion: namely, $\mathcal{G}_{\tau} \neq T$. This difference stems from the constraint equations imposed by the torsion perturbations.

Finally, let us bring (\ref{eqn ql0 scalar}) to a more useful form by using the equation for the  Lagrange multiplier $B$ ($\alpha=\frac{1}{a}\frac{ \mathcal{G}_{\tau}}{ \, \Theta}\dot{\psi} $) in $\mathcal{S}_{s} $. Thus, we obtain a single scalar mode
\begin{eqnarray}
\mathcal{S}_{s}= \int\, \textrm{d}\eta\,\textrm{d}^3x \,a^4\,\left(\frac{1}{a^2}\, \mathcal{G}_{\mathcal{S}}\,\dot{\psi}^2-\, \frac{1}{a^2}\, \mathcal{F}_{\mathcal{S}}\,(\partial_i \psi)^2\right)\,,\label{eqn ql finalStep}
\end{eqnarray}
where
\begin{eqnarray}
\begin{array}{cc}
\mathcal{G}_{\mathcal{S}}= 3\, \mathcal{G}_{\tau}+\frac{\mathcal{G}_{\tau}^2\,\Sigma}{\Theta^2} \label{eqn Gs}\,, \,&
\mathcal{F}_{\mathcal{S}}= \frac{1}{a^2}\frac{\text{d}}{\text{d}\eta}\left(\frac{a\, \mathcal{G}_{\tau}\,T}{\Theta}\right) -\mathcal{F}_{\tau}\,. \label{eqn Fs}
\end{array}
\end{eqnarray}

The no-go theorem follows a similar reasoning as in \cite{Rubakov:2016zah} in relation to wormholes, or as initially proved for a subclass of generalized Galileons in \cite{Libanov:2016kfc} and then extended to the full Horndeski action in \cite{Kobayashi:2016xpl}: 

\paragraph{No-Go for nonsingular, all-time stable and subluminal solutions:} For Galileons on a spacetime with torsion (\ref{eqn HC234c0action}) the following assumptions for a first order perturbative expansion about FLRW are mutually inconsistent
\begin{enumerate}[I)]
\item{Nonsingular cosmology: namely, there is a lower bound on the scale factor $a(\eta)>b_1>0$.}\label{eqn nonsingular cosmology}
\item{The graviton and the scalar mode are not ghosts and they suffer no gradient instabilities: 

$\mathcal{G}_{\tau}>0 \,, \mathcal{F}_{\tau}>0 \,,  \mathcal{F}_{\mathcal{S}}>0 \,,  \mathcal{G}_{\mathcal{S}}>0 $.}\label{eqn stability conditions}
\item{The graviton is always sub/ luminal: $(c_g)^2\leq 1$}\label{eqn subluminality condition} 
\item{There is a lower bound $\mathcal{F}_{\tau}(\eta)>b_2>0$ as $\eta\rightarrow \pm\infty$ (No "Strong gravity" at linear order \cite{Kobayashi:2016xpl,Ageeva:2021yik}).}\label{eqn no strong gravity}
\item{$\Theta$ vanishes at most a finite amount of times (To cover generic theories not defined by the equation $\Theta \equiv 0$ \cite{Mironov:2022quk}).}\label{eqn general Theta}
\end{enumerate}

The argument: It is key to notice that (\ref{eqn nonsingular cosmology})-(\ref{eqn subluminality condition}) imply
\begin{eqnarray}
N=: \frac{a\, \mathcal{G}_{\tau}\, \mathcal{F}_\tau\,(c_g^2-2)}{\Theta}\neq 0\,,\label{eqn ineq N}
\end{eqnarray} 
because by (\ref{eqn nonsingular cosmology}-\ref{eqn stability conditions}) $\Theta$ is a regular (finite) function of $H,\,$ $\phi$.

Let us integrate the third inequality in (\ref{eqn stability conditions}). Using (\ref{eqn Fs}) 
\begin{gather}
\Delta N=N_f-N_i> I(\eta_i,\,\eta_f)\,, \label{eqn ineq important}\\
I(\eta_i,\,\eta_f)=\int_{\eta_i}^{\eta_f}\text{d}\eta\,a^2\, \mathcal{F}_{\tau}\,, \nonumber
\end{gather}
where $N_f$ and $N_i$ are the values of $N$ at some (conformal) times $\eta_f$ and $\eta_i$ respectively. Now, by (\ref{eqn nonsingular cosmology}), (\ref{eqn stability conditions}) and (\ref{eqn no strong gravity})

\begin{enumerate}[A)]
\item{$\frac{d\, N}{d\,\eta}> a^2\, \mathcal{F}_{\tau} >0$,
\begin{itemize} 
\item{ defining $I(\eta_i) :=I(\eta_i,\,\eta_f)\vert_{\eta_f}$ and $I(\eta_f) :=I(\eta_i,\,\eta_f)\vert_{\eta_i} $ we notice that they are {\it positive and growing} functions of  $\eta_i$ and $\eta_f$, for $\eta_f$ and $\eta_i$ fixed, respectively. $I(\eta_i)$ and $I(\eta_f)$ are differentiable and hence continuous because $a^2\,\mathcal{F}_{\tau}$ is continuous,}
\item{$N$ is monotonous increasing and hence, denoting with $\eta_z$ any zero of $\Theta$, then $N(\eta)\rightarrow \infty$ as $\eta\rightarrow\eta_{z}^-$ ($\eta$ approaches $\eta_{z}$  by the {\it left}) and $N(\eta)\rightarrow -\infty$  as $\eta\rightarrow\eta_{z}^+$,}
\end{itemize}}\label{eqn conclusion 1}
\item{$\Delta N>0$,}
\item{$I(\eta_i) $ and $I(\eta_f)$ are not convergent as $\eta_i\rightarrow -\infty$, $\eta_f\rightarrow \infty$, respectively.}\label{eqn conclusion 3}
\end{enumerate}
Now, by (\ref{eqn general Theta}) $N$ is finite almost everywhere, thus let us take a fixed value $-\infty<N_i<0 $ at some fixed time $\eta_i$. By (\ref{eqn ineq N}) follows $N_f(\eta_f)<0$ (Without loss of generality we can safely assume that there is no $\eta_z$ (a zero of $\Theta$) such that $\eta_i<\eta_z<\eta_f$ \footnote{Indeed, if $\Theta$ has one zero $\eta_z$ such that $\eta_i<\eta_z<\eta_f $, then by (\ref{eqn conclusion 1}) $N$ must be {\bf positive} arbitrarily close {\bf by the left of} $\eta_z$. But our fixed value $N_i(\eta_i)<0$ means that $N(\eta)$ must have already vanished at some time $\eta$, with $\eta_i<\eta<\eta_z<\eta_f$ (because $N(\eta)$ is continuous for $\eta_i<\eta<\eta_z$), {\bf thus already violating} (\ref{eqn ineq N}). This clearly extends to any number of zeros. Thus, provided our starting point $-\infty<N_i(\eta_i)<0 $, we can exclude the case of any $\eta_z$ in the time interval $(\eta_i,\,\eta_f)$. \label{footnote1}}). Then $\Delta N $ is also bounded by above as $\vert N_i\vert>\Delta N=\vert N_i\vert-\vert N_f\vert >0$. Now, by \ref{eqn conclusion 1}), \ref{eqn conclusion 3}) $I(\eta_f)$ not only grows with $\eta_f$ but it is also unbounded from above, then there exists late enough in the evolution a critical time $\eta_c$ such that if $\eta_f>\eta_c$, $I(\eta_f)>\vert N_i\vert> \Delta N $ for every fixed value $N_i$. This violates (\ref{eqn ineq important}) and so we must have $N(\eta)>0$. However, by a similar argument, fixing a value $\infty>N_f>0$ at some $\eta_f$, necessarily $N_i(\eta_i)>0$ \footnote{By a similar argument as in \footref{footnote1}, provided our starting point $\infty>N_f>0$  we can exclude w.l.o.g. any $\eta_z$ in the time interval $(\eta_i,\,\eta_f)$.}, then $N_f>\Delta N= N_f-N_i >0$, and by \ref{eqn conclusion 1}), \ref{eqn conclusion 3}) $I(\eta_i)$ is unbounded from above and there exists $\eta_c$ early enough, such that if $\eta_i<\eta_c$, $I(\eta_i)>N_f> \Delta N $. Thus, eventually in the evolution (\ref{eqn ineq important}) does not hold.

In fact, (\ref{eqn subluminality condition}) can be relaxed to $c_g^2<2$ and the argument still holds. But, an almost everywhere subluminal graviton turning to $c_g^2\geq 2$ during an arbitrarily short interval is enough to avoid this No-Go argument, as we show below. Clearly, a minimal example needs $G_{4,X}\neq 0$.

\section{Everywhere Stable Bouncing Cosmology in Galileons with Torsion}\label{sec bounce}

A sufficient assumption to by-pass the no go theorem and obtain all-time stable solutions is, for instance, a period of  {\it non zero} width of superluminality of the graviton at some point in the evolution and by at least an amount $c_g\geq \,\sqrt{2}\,c $ ($c=1$ in our units). Thus, in principle, the width $\tau_s$ of an arguably unphysical superluminal phase centered at a time $\eta_s$ can be {\it arbitrarily short} and unrelated to the width $\tau_b$ and time of ocurrence of the physically relevant bounce phase centered at a time $\eta_b$. 

As a proof of principle let us show with a toy model that one can achieve stability of the FLRW bouncing background against linear order perturbations even when there is no relation between the time scales associated with the superluminal phase, necessary to avoid the no-go theorem, and the bounce phase. We assume that $\eta_s,\,\eta_b$ are finite and without loss of generality $\eta_s<\eta_b=0$. Furthermore, we demand for our model that the solution reduces in the asymptotic past and future to a solution that one could also obtain from conventional Einstein's gravity with a {\it luminal } graviton, minimally coupled to a massless scalar, and such that the torsion background is asymptotically vanishing. More precisely, we consider that $a(\eta)$ is positive and bounded from below, the bounce happening at the minimum $a(\eta_b)$, and for the latter asymptotics we require that the leading terms of the Lagrangian functions in (\ref{eqn HC234c0action}) and the torsion background behave as follows as $\eta\rightarrow \pm \infty $
\begin{eqnarray}
\begin{array}{cc}
G_2(\phi,X)\rightarrow\,\frac{1}{2\,a^2}\,\dot{\xi}^2\,, \label{eqn G2 asymptotics} & G_4(\phi,X) \rightarrow\,\frac{1}{2}\,, \label{eqn G4 asymptotics} \\
{}&{}\\
G_3(\phi,X) \rightarrow\,0\,, \label{eqn G3 asymptotics} & x(\eta)\, \rightarrow\,0\,. \label{eqn x asymptotics}
\end{array}
\end{eqnarray}
where $\xi $ is some invertible function of the Horndeski scalar $\phi$, and we choose $M_{pl}^2/8\pi=1$.
\subsection{Construction of the model}

\begin{figure}
\centering
  \includegraphics[width=1\linewidth]{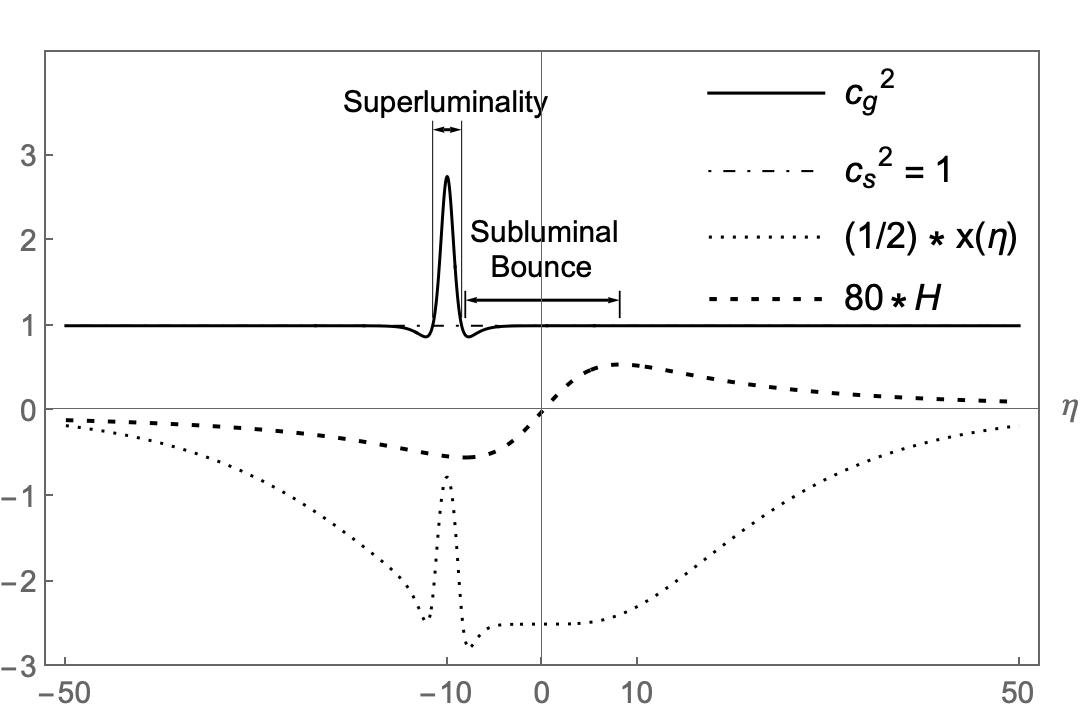}
 \caption{Hubble parameter for a bounce at $\eta_b=0$ with $\tau_b=10$. Speed of sound for the scalar mode $c_s^2$. Speed of the graviton $c_g^2$ with {\it short} superluminality phase ($\tau_s<<\tau_b$) happening at $\eta_s=-10$ before the bounce (For convenience displaying the graphs we choose here $\tau_b=10\,\tau_s$). The graviton quickly becomes subluminal around $\eta_s$ and approaches luminality from below in the past, and during the bounce phase and future. Torsion background $x(\eta)$ exponentially vanishing in the asymptotic past and future.} \label{fig backgroundsHxSpeed}
\end{figure}

\paragraph{Procedure:}
The following Ansatz for the Lagrangian functions has enough structure such that we can satisfy the asymptotic conditions (\ref{eqn G2 asymptotics}), besides demanding the stability {\it at all times}  (\ref{eqn stability conditions}), while simultaneously solving all the equations of motion:
\begin{eqnarray}
G_2(\phi,X)&=&g_{20}(\phi)\,+\, g_{21}(\phi)\,X\,+\, g_{22}(\phi)\,X^2\,, \label{eqn G2 ansatz}\\
G_3(\phi,X)&=&g_{30}(\phi)\,+\, g_{31}(\phi)\,X\,, \label{eqn G3 ansatz}\\
G_4(\phi,X)&=&\frac{1}{2}\,+\,g_{40}(\phi)\,+\, g_{41}(\phi)\,X\,. \label{eqn G4 ansatz}
\end{eqnarray}
We reconstruct the $7$ unknown Lagrangian functions in (\ref{eqn G2 ansatz})-(\ref{eqn G4 ansatz}), namely, $g_{20},\, g_{21},\, g_{22},\, g_{30},\, g_{31},\, g_{40} $ and $g_{41} $ stating first some solutions satisfying our requirements for $a(\eta)$ and the asymptotics (\ref{eqn G2 asymptotics}), and then we work backwards to find the Lagrangian functions whose dynamics correspond to the latter. We proceed as follows: without loss of generality we choose a model with a solution for the Hubble parameter, shown in Figure \ref{fig backgroundsHxSpeed}, and the background of the Horndeski scalar field as
\begin{eqnarray}
\begin{array}{ccc}
a\,=\, (\tau_b^2+\eta^2)^{\frac{1}{4}}\,, & H\,=\,\frac{\dot{a}}{a^2}\,=\, \frac{\eta}{2\,(\tau_b^2+\eta^2)^{\frac{5}{4}}}\,, &\, \phi=\eta \,,\label{eqn sol backgrounds}
\end{array}
\end{eqnarray}
such that our definition of bounce is satisfied. $\tau_b>0$ fixes the maximum of $H$ and the width of the bounce phase as the length of the domain where $\dot{H}(\eta)>0$ around $\eta_b$. With this solution $X=1/(2\, (\tau_b^2+\eta^2)^{\frac{1}{2}})$. Now we solve the unknown Lagrangian functions in (\ref{eqn G2 ansatz})-(\ref{eqn G4 ansatz}):

\paragraph{Solving for $g_{40}$ and $g_{41}$ provided $\mathcal{G}_{\tau}>0$, $\mathcal{F}_{\tau}>0$ for all time, $T$ vanishing at least once at some critical time and $G_4$ asymptotics (\ref{eqn G4 asymptotics}): }

\begin{figure}
\centering
  \includegraphics[width=1\linewidth]{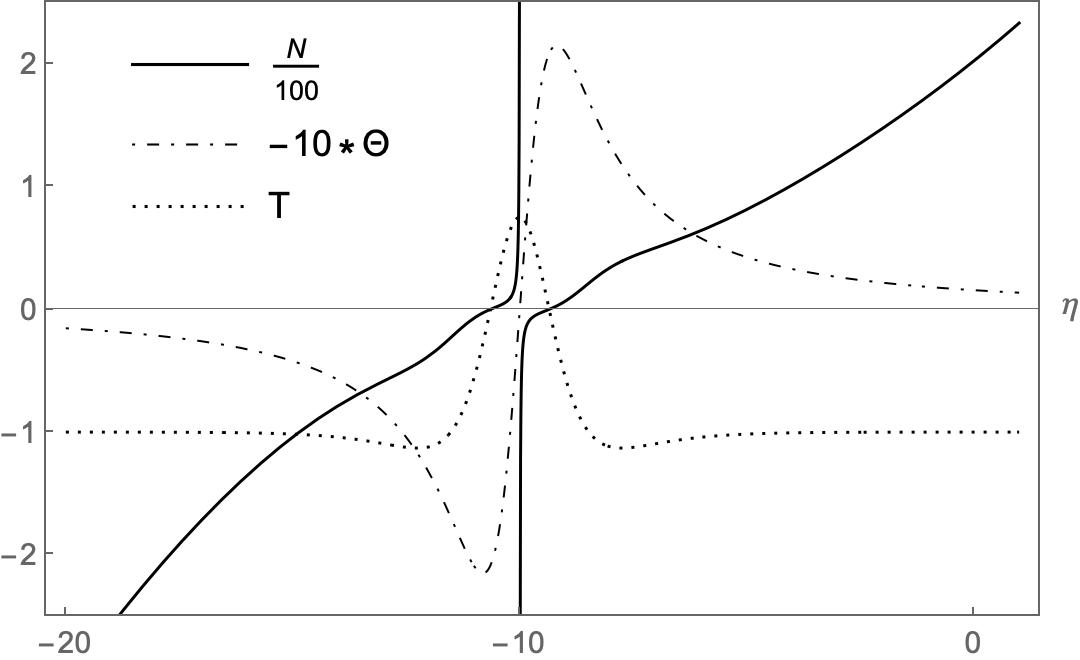}
 \caption{By-passing the no-go theorem: this choice for $T$ (\ref{eqn controled superluminality}) does not satisfy the all-time negativity condition, which critically means that the graviton is superluminal during a brief stage of evolution, around $\eta_s=-10$ as shown in Figure \ref{fig backgroundsHxSpeed}, and that the function $N$ in equation (\ref{eqn ineq N}) vanishes. Hence, the no-go theorem does not hold and we can build all-time stable solutions. $(\tau_b=10,\,\tau_s=1,\, \eta_b=0,\, \eta_s=-10)$} \label{fig thetaNT}
\end{figure}

$\mathcal{G}_{\tau} $, $\mathcal{F}_{\tau} $ and the critical function $T$ (\ref{eqn T}) depend only on $G_4$. Hence, we can solve for $g_{40}$ and $g_{41}$ from two algebraic equations in these variables
\begin{eqnarray} 
&&\mathcal{F}_{\tau}(g_{40},\,g_{41})=1\,, \label{eqn G4 asymptotics practical} 
\end{eqnarray}
\begin{equation}
\resizebox{.85\hsize}{!}{$T(g_{40},\,g_{41})= -1-\frac{5}{4}\text{Sech}\left(\frac{\eta-\eta_s}{\,\tau_s}\right)\,+\,3\,\text{Sech}\left(\frac{\eta-\eta_s}{\tau_s}\right)^2$} \label{eqn controled superluminality}
\end{equation}
Equation (\ref{eqn G4 asymptotics practical}) is a simple {\it choice} to realize the desired asymptotics of $G_4$ (\ref{eqn G4 asymptotics}) and to obviously satisfy $\mathcal{F}_{\tau}>0$. The equation (\ref{eqn controled superluminality}) is an explicit {\it choice} to violate the subluminal graviton assumption at least during a short time $\tau_s<<\tau_b$, which allows to bypass the no-go theorem, as is shown in Figure \ref{fig thetaNT}. The solutions for $g_{40}$ and $g_{41}$ from the system of equations (\ref{eqn G4 asymptotics practical}), (\ref{eqn controled superluminality}) are straightforward and everywhere nonsingular. Their graphs are shown in Figure \ref{fig G4s} in the Appendix \ref{sec figures}. It can be readily verified that these solutions also imply $\mathcal{G}_{\tau}>0$. They can be written at leading order as $\eta\rightarrow \pm \infty$ in the form
\begin{eqnarray}
g_{40}=-g_{41}\,X=\frac{5}{8}\,e^{\mp\frac{(\eta-\eta_s)}{\tau_s}}\,. \label{eqn G4s asymptotic solution}
\end{eqnarray}

\paragraph{Solving for $g_{30}$ and $g_{31}$ provided $\mathcal{F}_{\mathcal{S}}>0$, torsionless and $G_3$ asymptotics (\ref{eqn x asymptotics}):} $\mathcal{F}_{\mathcal{S}}$ and the torsion background ($x$) depend on $G_4$, which is now fully fixed, and on $G_3$.  Hence, we can solve for $g_{30}$ and $g_{31}$ from two equations that are algebraic in these functions
\begin{eqnarray}
\begin{array}{cc}
G_3(g_{30},\,g_{31})= \text{Sech}\left(\frac{\eta}{\tau_b}\right)\,, \label{eqn G3 asymptotics practical} & \Theta(g_{30},\,g_{31})=-H_s \label{eqn Theta practical}\,,
\end{array}
\end{eqnarray}
where
\begin{eqnarray}
H_s&=&\frac{\eta-\eta_s\,S}{2\,(\tau_b^2\,(1-S)+\tau_s^2\,S\,+(\eta-\eta_s\,S)^2)^{\frac{5}{4}}}\,,
\end{eqnarray}
and $S$ is a step function which we specify below.

The {\it choice} of $G_3$ centered at the bounce $\eta_b=0$ (\ref{eqn G3 asymptotics practical}) vanishes exponentially fast in the asymptotic past and future, which satisfies (\ref{eqn G3 asymptotics}). On the other hand, we choose an equation for $\Theta$ because it is the only free function in $\mathcal{F}_{\mathcal{S}}$. It has to satisfy the two remaining conditions:  $(i)$ it must be in accordance to the required asymptotics (\ref{eqn x asymptotics}) and $(ii)$ it must render $\mathcal{F}_{\mathcal{S}}$ positive everywhere. For $(i)$ a close inspection of $\Theta$ in terms of the Ansatz (\ref{eqn G2 ansatz})-(\ref{eqn G4 ansatz}) reveals that in order to recover a standard scalar minimally coupled to Einstein's gravity (\ref{eqn x asymptotics}), then one needs $\Theta \overset{\eta\rightarrow \pm \infty}{\longrightarrow} -H$ \footnote{Similar as in the torsionless case, this choice introduces a well-known removable singularity in the unitary gauge known as $\gamma$-crossing  that can be seen to be harmless for the regularity of perturbations as in \cite{Mironov:2018oec}.}. Hence, in the equation (\ref{eqn Theta practical}) the {\it first requirement} for the step function $S$ is that it suppresses the factors $\tau_s$ and $\eta_s$ in $H_s$ fast enough such that in the asymptotic past an future we recover asymptotics of our Hubble parameter (\ref{eqn sol backgrounds}) {\it at the necessary order} in $\eta$ \footnote{In particular, it is not a trivial fact that one must choose the step function $S$ such that the limit $\Theta=-H_s \rightarrow -H$ as $\eta\rightarrow \pm \infty$ is satisfied at {\bf more} than leading order in $\eta$, in order to meet the required asymptotics of the Lagrangian functions (\ref{eqn G2 asymptotics}) {\bf only} at the leading order.} that satisfies the asymptotics (\ref{eqn G2 asymptotics}). On the other hand, $(ii)$ can be easily satisfied with a step $S$ that is nearly $1$ in a domain of finite length $(\eta_s-\delta,\,\eta_b+\delta)$ where $\infty>\delta>0$, for a large enough $\delta$ depending on the parameters $\tau_b,\,\tau_s,\,\eta_b,\,\eta_s$. As a proof of principle we choose
\begin{eqnarray}
S=\text{Sech}\left(\frac{\tau_s}{\tau_b}\frac{(\eta-\eta_s)}{\,\eta_s}\right)\,,
\end{eqnarray}
which satisfies $(i)$ and $\mathcal{F}_{\mathcal{S}}>0$ everywhere. Suffices to say for this toy model that this choice meets the requirements, as shown in Figure \ref{fig thetaNT}, for instance for the parameters of the bounce $\eta_b=0\,,\,\tau_b=10$ and of the earlier and {\it shorter} superluminal phase $\eta_s=-10\,,\,\tau_s=\tau_b/10$. 

The solutions for $g_{30}$ and $g_{31}$ obtained from (\ref{eqn G3 asymptotics practical}) are everywhere regular and their graphs are shown in Figure \ref{fig G3s}. They can be written at leading order as $\eta\rightarrow \pm \infty$ in the form
\begin{eqnarray}
g_{30}=-g_{31}\,X=\frac{3}{2}\frac{\,\eta_s}{\eta^2} e^{\mp \frac{\tau_s}{\tau_b}\frac{(\eta-\eta_s)}{\,\vert\eta_s\vert}} \,,\label{eqn G3s asymptotic solution}
\end{eqnarray}
where we used $\eta_s<0$, and the torsion background can be written at leading order as $\eta\rightarrow \pm \infty$, as
\begin{eqnarray}
x=\mp \eta\,e^{\mp\frac{\eta}{\tau_b}}\,.\label{eqn solution torsion background}
\end{eqnarray}
\paragraph{Solving for $g_{20},\, g_{21}$ and $g_{22}$ provided $\mathcal{G}_s>0$, always sub/ luminal scalar and the E-L equations for the background fields:} 
$\mathcal{G}_s $ and the Euler-Lagrange equations for the background fields $\mathcal{E}_{g_{00}}=0$, $\mathcal{E}_{g_{ij}}=0$ depend on $G_3,\, G_4$, which are fully fixed, and on $G_2$. Hence, we can solve for the lagrangian functions $g_{20}(\phi),\, g_{21}(\phi)$ and $g_{22}(\phi)$ from the following system of three equations, which again, is algebraic and linear in these functions
\begin{eqnarray}
\begin{array}{ccc}
\mathcal{G}_{\mathcal{S}}=\mathcal{F}_{\mathcal{S}} \label{eqn Gs practical}\,,& \mathcal{E}_{g_{00}}=0 \label{eqn g00 practical}\,, & \mathcal{E}_{g_{11}}= \mathcal{E}_{g_{22}}= \mathcal{E}_{g_{33}}=0 \label{eqn gii practical}\,.
\end{array}
\end{eqnarray}
Because the lagrangian functions $g_{30}(\phi)$ and $g_{31}(\phi)$ are such that $\mathcal{F}_{\mathcal{S}}>0$, the choice of equation (\ref{eqn Gs practical}) is one possibility to simultaneously satisfy a non ghost scalar $\mathcal{G}_{\mathcal{S}}>0$ and luminality for the scalar mode $c_s^2= \mathcal{F}_{\mathcal{S}}/\mathcal{G}_{\mathcal{S}}=1 $. We choose luminality for no other reason than simplicity, although a subluminal choice is safer and better suited in many other cases. The unique solutions for $g_{20}(\phi),\, g_{21}(\phi)$ and $g_{22}(\phi)$ obtained from (\ref{eqn Gs practical}) are non singular everywhere and their graphs are shown in Figure \ref{fig G2s around superluminality}. They can be written at leading order as $\eta\rightarrow \pm \infty$ as
\begin{eqnarray}
\begin{array}{cc}
g_{20}=-\frac{\tau_b^2}{2}\,\left(\pm\eta\right)^{-5}\,, \label{eqn G2s asymptotic solution 1} & g_{21}\,X=\frac{3}{4}\,\left(\pm\eta\right)^{-3}\,, \label{eqn G2s asymptotic solution 1}
\end{array}
\end{eqnarray}
\begin{eqnarray}
g_{22}\,X^2&=&\mp\frac{3}{4}(\pm\eta)^{-3} \frac{\tau_s}{\tau_b} e^{\mp\frac{\tau_s}{\tau_b}\frac{(\eta-\eta_s)}{\,\vert\eta_s\vert}}\,.\label{eqn G2s asymptotic solution 2}
\end{eqnarray}
Let us note that because $G_2,\, G_3,\, G_4$ are such that the E-L equations (\ref{eqn g00 practical}) and their derivatives are satisfied, then the bouncing solution  (\ref{eqn sol backgrounds}) is the correct one for the model with the Lagrangian functions that we just solved. Furthermore, the remaining E-L equation for the background scalar $(\mathcal{E}_{\phi}=0)$ is implied by the others because of gauge invariance, as can be readily verified, which certifies that the solution $\phi=\eta$ in (\ref{eqn sol backgrounds}) is also the correct one for the model with the lagrangian functions just built.

\subsection{Asymptotic Lagrangian: recovering GR}\label{sec asymptotics}
The leading order expressions of the Lagrangian functions as $\eta\rightarrow \pm \infty$ (\ref{eqn G4s asymptotic solution}), (\ref{eqn G3s asymptotic solution}), (\ref{eqn G2s asymptotic solution 1}), (\ref{eqn G2s asymptotic solution 2}) in the Ansatz (\ref{eqn G2 ansatz})-(\ref{eqn G4 ansatz}) tell that to the leading order the only non vanishing Lagrangian functions are $G_4=\frac{1}{2}$ and $G_2$. Hence, with the solution $\phi=\eta$ and the leading order expression $X=1/(2\,\eta)$, considering only the leading contributions to $G_2$, namely $ g_{21}\,X$, we can identify at leading order the corresponding action to (\ref{eqn HC234c0action}) in the asymptotic past and future
\begin{eqnarray}
S^{\infty}=\frac{1}{2}\int\text{d}^4 x \sqrt{-g} \left(R- \partial_\mu\xi\, \partial^\mu\xi\right)\,\label{eqn asymptotic lagrangian}
\end{eqnarray}
for a massless scalar field $\xi=\sqrt{\frac{3}{2}}\,\text{ln}(\phi)$ minimally coupled to Einstein's gravity and with vanishing torsion background (\ref{eqn solution torsion background}). Indeed, one can check that the field and Friedmann equations derived from (\ref{eqn asymptotic lagrangian}) are satisfied by the leading order contributions as $\eta\rightarrow \pm\infty$ of the solutions that we started with in (\ref{eqn sol backgrounds}): namely, $a=\eta^{\frac{1}{2}},\, H=\frac{1}{2}\eta^{-\frac{3}{2}},\, \phi=\eta,\, \xi=\sqrt{\frac{3}{2}}\,\text{ln}(\eta) $ solve
\begin{eqnarray}
\begin{array}{cc}
\ddot{\xi}+2\,a\,H\,\dot{\xi}=0\,, & \dot{\xi}^2-6\,a^2\,H^2=0\,.
\end{array}
\end{eqnarray}


\section{Conclusion}\label{sec conclusions} \label{sec gamma crossing}

We first extended the no-go argument of \cite{Libanov:2016kfc, Kobayashi:2016xpl} to Galileons on a spacetime with torsion (Horndeski-Cartan) (\ref{eqn HC234c0action}). We showed that in generic models it is not possible to obtain a nonsingular FLRW cosmology that is always free of gradient instabilities against the scalar perturbation and an eternally sub/ luminal graviton. 

Then, we highlighted that unlike in the torsionless theory, where instabilities happen with certainty at some time in the entire evolution \cite{Libanov:2016kfc, Kobayashi:2016xpl}, a spacetime with torsion can support all-time {\it linearly} stable nonsingular solutions in Galileons if there exists at an arbitrary time a superluminal phase for the graviton and by at least an amount $c_g\geq \,\sqrt{2}\,c $. This unphysical phase could formally happen as a  deep UV inconsistency, namely, arbitrarily short and unrelated to the physically relevant length scales that are pertinent to these models, such as time and much longer width of a bounce. Besides, this pathology in the classical solutions may still be informative raising the question about the possibility of Lorentz invariant UV completions \cite{Adams:2006sv, Creminelli:2022onn} and whether causal paradoxes arise \cite{nicolis2009galileon, Evslin:2011vh, Dubovsky:2005xd}. Finally, we showed a bouncing cosmology that is always stable, where a short superluminal phase happens before the bounce and that transits to Einstein's gravity coupled to a massless scalar and with vanishing torsion in the asymptotic past and future.

At least in what concerns the stability and speed of solutions, this shows that Horndeski-Cartan theory is fundamentally different to Horndeski on a torsionless geometry, in contrast to {\it e.g.} the equivalence of Einstein-Cartan and GR.

\section*{Acknowledgements}

The work of S.M. on Sec. I of this paper has been supported by Russian Science Foundation grant 19-12-00393, while the part of work on Sec. II has been supported by the Foundation for the Advancement of Theoretical Physics and Mathematics “BASIS”.


\section{Appendix}

\subsection{Derivation of the quadratic action:}\label{sec notation}
To derive expressions (\ref{eqn qlf tensor}), (\ref{eqn ql0 scalar}) we follow the notation and detailed procedure in \cite{Mironov:2023kzt}. Briefly: we consider the perturbed metric $\textrm{d}s^2=\left(\eta_{\mu\nu}+\delta g_{\mu\nu}\right)\textrm{d}x^\mu\, \textrm{d}x^\nu $ where $\eta_{\mu\nu}\textrm{d}x^\mu\, \textrm{d}x^\nu= a^2(\eta)\left(-\textrm{d}\eta^2+\delta_{ij}\, \textrm{d}x^i \,\textrm{d}x^j \right)$ is a spatially flat FLRW background metric, $\eta$ is conformal time, and we denote spatial indices with latin letters such as $i=1,2,3$ and space-time indices with greek letters, such as $\mu=0,1,2,3$. The metric perturbation is written as 
\begin{dmath}
\delta g_{\mu\nu}\,\textrm{d}x^\mu\, \textrm{d}x^\nu
\,\text{$=$}\, a^2(\eta)\left(-2\,\alpha\,\textrm{d}\eta^2+2\left(\partial_i B+S_i\right) \textrm{d}\eta \, \textrm{d}x^i+\left(-2\,\psi\, \delta_{ij}+2\,\partial_i\partial_j E+\partial_i F_j+\partial_j F_i+2\,h_{ij}\right) \textrm{d}x^i \, \textrm{d}x^j \right)\,,
\end{dmath}
with $\alpha,\, B,\, \psi,\, E$  scalar perturbations, $S_i,\, F_i$  transverse vector perturbations, and $h_{ij}$, a symmetric, traceless and transverse tensor perturbation. 

On the other hand we consider the perturbed Galileon field $(\phi(x))$ as $\phi(\eta)+\Pi(\eta,\vec{x})$ in the linearized expressions, where in this context $\phi(\eta) $ is the background field. 

To write explicit torsion in (\ref{eqn HC234c0action}), we denote torsionful and metric compatible covariant derivative on any vector $V^\mu$ as $\tilde{\nabla}_\mu V^\nu=\partial_\mu V^\nu+\tilde{\Gamma}^{\nu}_{\mu\lambda}V^{\lambda}$, where the non-symmetric torsionful connection can be expressed in terms of the usual GR Christoffel connection $\Gamma^{\rho}_{\mu\nu}=\frac{1}{2}g^{\rho\sigma}\left(\partial_\mu g_{\nu\sigma}+\partial_\nu g_{\mu\sigma}-\partial_\sigma g_{\mu\nu}\right) $, as $\tilde{\Gamma}^{\rho}_{\mu\nu}=\Gamma^{\rho}_{\mu\nu}-K^{\rho}{}_{\mu\nu}$ (Namely, we introduce torsion in the second order formalism). $K^{\rho}{}_{\mu\nu} $ is named contortion tensor and with our convention of torsionful covariant derivatives it is antisymmetric in the first and third indices, $K_{\mu\nu\sigma}= -K_{\sigma\nu\mu} $, such that it  has $24$ independent components.  

The perturbed contortion tensor $K_{\mu\nu\sigma}= {}^{0}K_{\mu\nu\sigma}(\eta)+ \delta K_{\mu\nu\sigma}(\eta,\vec{x}) $ has only two non vanishing background contributions on an isotropic and homogeneous spacetime, namely: ${}^{0}K_{0\,j\,k}=\,x(\eta)\,\delta_{jk}$, and ${}^{0}K_{i\,j\,k}=\,y(\eta)\epsilon_{i\,j\,k}$. For the spacetime dependent perturbation $\delta K_{\mu\nu\sigma}(\eta,\vec{x}) $ the $24$ independent components can be written in terms of irreducible components under small rotation group as: eight scalars denoted as $C^{\scalebox{0.5}{(n)}} $ with $n=1, \dots , 8$, six (two-component) transverse vectors and two (two-component) traceless, symmetric, transverse tensors $T^{\scalebox{0.5}{(1)}}_{ij},\, T^{\scalebox{0.5}{(2)}}_{ij}$. An explicit decomposition is given, for instance, in section $II.B$ in \cite{Mironov:2023kzt}.

The 4 background fields $\phi,\,a,\,x,\,y$ obey $5$ equations of which only $4$ are independent (due to gauge redundancy), which we denote as ${\mathcal{E}}_{f}=\partial \mathcal{L}/\partial f\,=\,0 $ for $f$ one of the following: $\phi,\, g_{00},\, g_{ij},\, {}^{0}K_{0\,j\,k},\, {}^{0}K_{i\,j\,k} $. In particular, $\mathcal{E}_{{}^{0}K_{ijk}} =-2\,\epsilon_{ijk}\,\,G_{4}\,y/a^6=0 $ sets $y(\eta)\equiv 0$ and $\mathcal{E}_{{}^{0}K_{0ij}}=0$ solves $x(\eta)$ in terms of $a,\,\phi$ 
\begin{eqnarray}
x(\eta)=-\frac{a^3\,\mathcal{G}_\tau\,(8\,H\,X\,G_{4,X}+\,a\,\dot{\phi}\,(G_3-2\,G_{4,\phi}))}{8\,G_4^2}\,.\label{eqn solx}
\end{eqnarray}

The quadratic action for the three tensor perturbations $h_{ij},\, T^{\scalebox{0.5}{(1)}}_{ij},\, T^{\scalebox{0.5}{(2)}}_{ij}$ is obtained as usual and implies $T^{\scalebox{0.5}{(2)}}_{ij}\equiv 0$ and
\begin{eqnarray}
T^{\scalebox{0.5}{(1)}}_{ij}=\frac{2\,a^2\,X\,G_{4,X}}{\,G_4+\,2\,X \, G_{4,X}}\dot{h}_{ij}-2\,x\,h_{ij}\,.
\end{eqnarray}
Using these equations back in the quadratic action gives (\ref{eqn qlf tensor}). Notice the difference in $\mathcal{G}_\tau$  between the graviton $h_{ij}$ in Galileons in a torsionful and a torsionless spacetime due to the non trivial $T^{\scalebox{0.5}{(1)}}_{ij} $.

Similarly, the part of the quadratic action relevant to the thirteen scalar perturbations $\Pi,\,\alpha,\, B,\, \psi,\, E,\, C^{\scalebox{0.5}{(n)}}$ (with $n=1, \dots , 8$) is written  as (\ref{eqn ql0 scalar}) after integrating out all of the eight non dynamical torsion perturbations $C^{\scalebox{0.5}{(n)}}$. This is simpler in the unitary gauge, where $\Pi=0$ and $E=0$, because one can recognize (See \cite{Mironov:2023kzt} for the theory with $c=0$) that there are five Lagrange multipliers $C^{\scalebox{0.5}{(1)}},\, C^{\scalebox{0.5}{(5)}},\, C^{\scalebox{0.5}{(7)}},\, C^{\scalebox{0.5}{(2)}},\, B $. The constraint equations imposed by the first three Lagrange multipliers imply the vanishing of  $C^{\scalebox{0.5}{(6)}},\, C^{\scalebox{0.5}{(4)}},\, C^{\scalebox{0.5}{(8)}}$ respectively and with the equation for $C^{\scalebox{0.5}{(2)}} $ one can express the only non trivial torsion scalar as
\begin{dmath}
C^{\scalebox{0.5}{(3)}}= -\frac{2\,a^2\,X\,G_{4,X}}{\,G_4+\,2\,X \, G_{4,X}}\dot{\psi}+2\,x\,\psi-\frac{a^3\,(2\,G_4\,H\,+\,\Theta)+a^2\,\dot{\phi}\,G_{4,\phi}}{2\,G_4}\alpha\,.\label{eqn c3}
\end{dmath}
Using (\ref{eqn c3}) in the quadratic action gives (\ref{eqn ql0 scalar}), where critically $\mathcal{G}_\tau\neq T$ as opposed to Galileons without torsion.

\subsection{Lagrangian functions}\label{sec figures}
The Lagrangian functions $g_{20},\, g_{21},\, g_{22},\, g_{30},\, g_{31},\, g_{40},$ and $g_{41} $ have an exact solution. We show below the graphs of these functions around the bounce at $\eta_b=0$ with width $\tau_b= 10$ and at the short superluminality phase at $\eta_s=-10$ with width $\tau_s<<\tau_b$ (we choose here $\tau_b=10\,\tau_s$ for convenience displaying the graphs):

\begin{figure}[h]
\centering
  \includegraphics[width=1\linewidth]{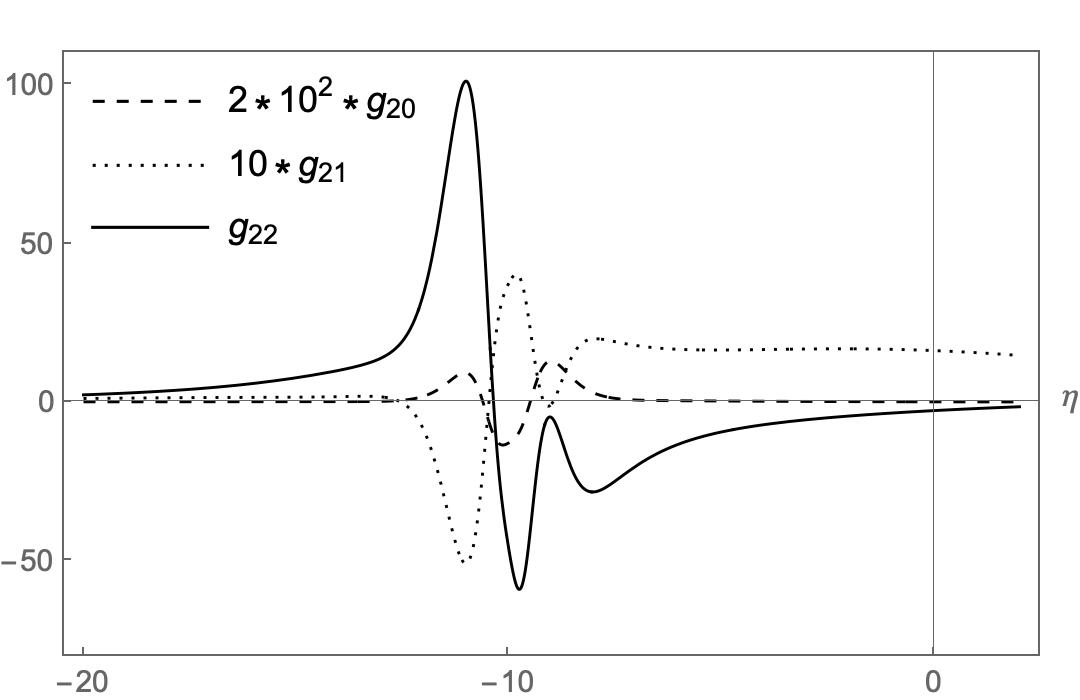}
 \caption{Everywhere regular Lagrangian functions $g_{20},\, g_{21}$ and $g_{22}$.} \label{fig G2s around superluminality}
\end{figure}

\begin{widetext}

\begin{figure}[h]
 \centering
\begin{subfigure}{0.4\textwidth}
\centering
  \includegraphics[width=1\linewidth]{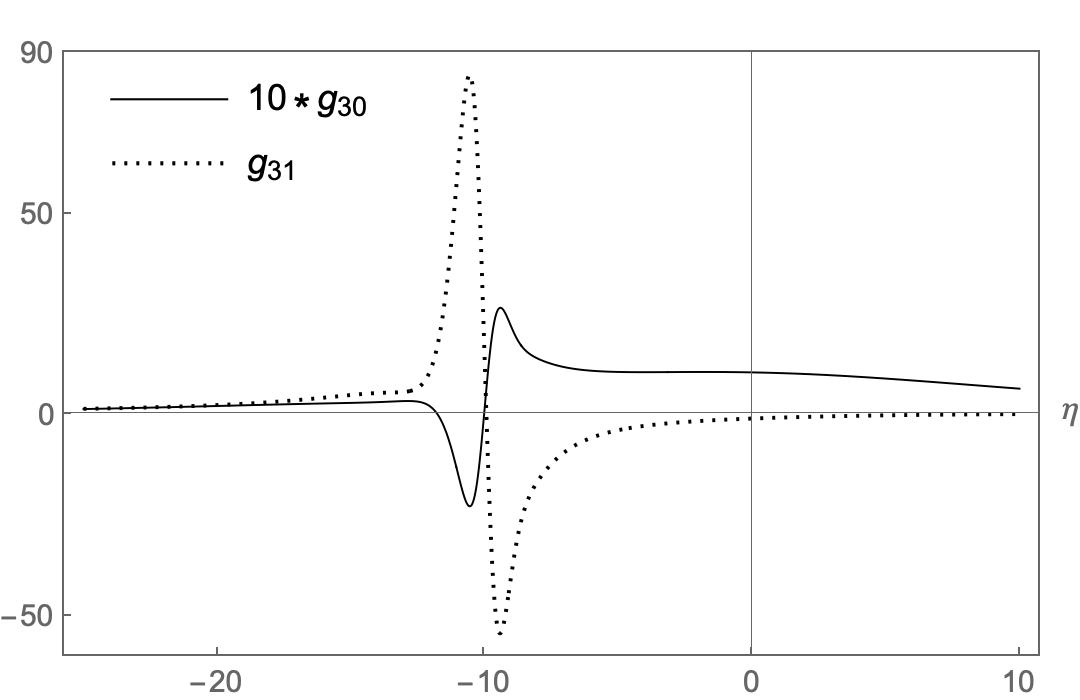}
 \caption{$g_{30}$ and $g_{31}$.} \label{fig G3s}
\end{subfigure}%
\begin{subfigure}{0.4\textwidth}
\centering
  \includegraphics[width=1\linewidth]{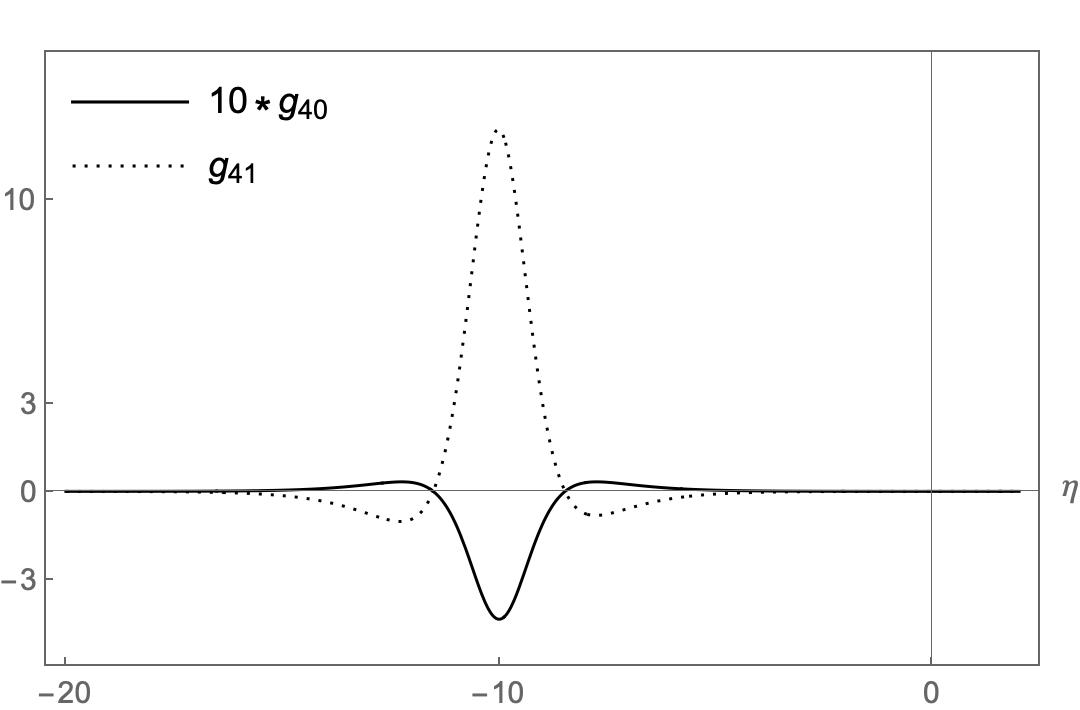}
 \caption{$g_{40}$ and $g_{41}$.} \label{fig G4s}
\end{subfigure}\caption{Everywhere regular Lagrangian functions.}
\end{figure}

\subsection{Coefficients in the final action for the scalar perturbations}\label{sec app Coeff final action}
The coefficients $\theta$ and $\sigma$ for the quadratic action (\ref{eqn ql0 scalar}) are
\begin{dmath}
\theta\,\text{=}\, -2 \,{a} \,{G_{4}} \,{H} (\,{G_{4}}^2 + 4 (2 \,{G_{4,X}}^2 -  \,{G_{4,XX}} \,{G_{4}}) \,{X}^2) -  (\,{G_{4,{\phi}}} (\,{G_{4}}^2 + 4 \,{G_{4,X}}^2 \,{X}^2 - 2 \,{G_{4}} \,{X} (\,{G_{4,X}} + 2 \,{G_{4,XX}} \,{X})) + \,{G_{4}} \,{X} (- (\,{G_{3,X}} - 2 \,{G_{4,{\phi}X}}) (\,{G_{4}} + 2 \,{G_{4,X}} \,{X}) + \,{G_{3}} (3 \,{G_{4,X}} + 2 \,{G_{4,XX}} \,{X}))) \,\dot{\phi}\,,
\end{dmath}
\begin{dmath}
\sigma\,\text{=} \,{a} \left(-24 \,{H}^2 (\,{G_{4}}^4 + 16 \,{G_{4,X}}^4 \,{X}^4 + 8 \,{G_{4,X}}^2 \,{G_{4}} \,{X}^3 (\,{G_{4,X}} - 2 \,{G_{4,XX}} \,{X}) + 2 \,{G_{4}}^2 \,{X}^2 (15 \,{G_{4,X}}^2 + 8 \,{G_{4,XX}}^2 \,{X}^2 \\
+ 4 \,{G_{4,X}} \,{X} (\,{G_{4,XX}} -  \,{G_{4,XXX}} \,{X})) -  \,{G_{4}}^3 \,{X} (\,{G_{4,X}} + 4 \,{X} (4 \,{G_{4,XX}} + \,{G_{4,XXX}} \,{X}))) + \,{X} (-3 \,{G_{3}}^2 (\,{G_{4}}^2 + 2 \,{X}^2 (5 \,{G_{4,X}}^2 + 8 \,{G_{4,XX}}^2 \,{X}^2 + 4 \,{G_{4,X}} \,{X} (\,{G_{4,XX}} -  \,{G_{4,XXX}} \,{X})) 
-  \,{G_{4}} \,{X} (11 \,{G_{4,X}} + 4 \,{X} (5 \,{G_{4,XX}}\\
 + \,{G_{4,XXX}} \,{X}))) + 6 \,{G_{3}} (- \,{X} (\,{G_{4}} + 2 \,{G_{4,X}} \,{X}) (2 (\,{G_{3,XX}} - 2 \,{G_{4,{\phi}XX}}) \,{X} (\,{G_{4}} + 2 \,{G_{4,X}} \,{X}) \\
+ \,{G_{3,X}} (5 \,{G_{4}} - 2 \,{X} (\,{G_{4,X}} + 4 \,{G_{4,XX}} \,{X})) + 2 \,{G_{4,{\phi}X}} (-5 \,{G_{4}} + 2 \,{X} (\,{G_{4,X}} + 4 \,{G_{4,XX}} \,{X}))) \\
+ 2 \,{G_{4,{\phi}}} (\,{G_{4}}^2 + 2 \,{X}^2 (5 \,{G_{4,X}}^2 + 8 \,{G_{4,XX}}^2 \,{X}^2 + 4 \,{G_{4,X}} \,{X} (\,{G_{4,XX}} -  \,{G_{4,XXX}} \,{X}))\\
 -  \,{G_{4}} \,{X} (11 \,{G_{4,X}} + 4 \,{X} (5 \,{G_{4,XX}} + \,{G_{4,XXX}} \,{X})))) + 4 ((\,{G_{4}} + 2 \,{G_{4,X}} \,{X})^2 (\,{G_{2,X}} (\,{G_{4}} + 2 \,{G_{4,X}} \,{X}) \\
- 2 \,{G_{3,{\phi}}} (\,{G_{4}} + 2 \,{G_{4,X}} \,{X})+ \,{X} (-3 (\,{G_{3,X}} - 2 \,{G_{4,{\phi}X}})^2 \,{X} + 2 (\,{G_{2,XX}} -  \,{G_{3,{\phi}X}}) (\,{G_{4}} + 2 \,{G_{4,X}} \,{X}))) \\
+ 3 \,{G_{4,{\phi}}} \,{X} (\,{G_{4}} + 2 \,{G_{4,X}} \,{X}) (2 (\,{G_{3,XX}} - 2 \,{G_{4,{\phi}XX}}) \,{X} (\,{G_{4}} + 2 \,{G_{4,X}} \,{X}) + \,{G_{3,X}} (5 \,{G_{4}} - 2 \,{X} (\,{G_{4,X}} + 4 \,{G_{4,XX}} \,{X})) \\
+ 2 \,{G_{4,{\phi}X}} (-5 \,{G_{4}} + 2 \,{X} (\,{G_{4,X}} + 4 \,{G_{4,XX}} \,{X}))) - 3 \,{G_{4,{\phi}}}^2 (\,{G_{4}}^2 + 2 \,{X}^2 (5 \,{G_{4,X}}^2 + 8 \,{G_{4,XX}}^2 \,{X}^2\\
 + 4 \,{G_{4,X}} \,{X} (\,{G_{4,XX}} -  \,{G_{4,XXX}} \,{X})) -  \,{G_{4}} \,{X} (11 \,{G_{4,X}} + 4 \,{X} (5 \,{G_{4,XX}} + \,{G_{4,XXX}} \,{X})))))\right)
 + 24 \,{H} \left(\,{X} ((\,{G_{4}} + 2 \,{G_{4,X}} \,{X}) ((2 \,{G_{3,X}} - 5 \,{G_{4,{\phi}X}}) \,{G_{4}}^2 + \,{G_{4}} (- (\,{G_{3,X}} + 2 \,{G_{4,{\phi}X}}) \,{G_{4,X}} + (\,{G_{3,XX}} - 2 \,{G_{4,{\phi}XX}}) \,{G_{4}}) \,{X} \\
+ 2 ((\,{G_{3,X}} - 4 \,{G_{4,{\phi}X}}) \,{G_{4,X}}^2 + ((\,{G_{3,XX}} - 2 \,{G_{4,{\phi}XX}}) \,{G_{4,X}} - 2 (\,{G_{3,X}} - 2 \,{G_{4,{\phi}X}}) \,{G_{4,XX}}) \,{G_{4}}) \,{X}^2) + \,{G_{3}} (-6 \,{G_{4,X}} \,{G_{4}}^2 + 3 \,{G_{4}} (\,{G_{4,X}}^2 - 3 \,{G_{4,XX}} \,{G_{4}}) \,{X} - 2 (3 \,{G_{4,X}}^3 - 2 \,{G_{4,X}} \,{G_{4,XX}} \,{G_{4}} + \,{G_{4,XXX}} \,{G_{4}}^2) \,{X}^2 \\
- 4 (\,{G_{4,X}}^2 \,{G_{4,XX}} - 2 \,{G_{4,XX}}^2 \,{G_{4}} + \,{G_{4,X}} \,{G_{4,XXX}} \,{G_{4}}) \,{X}^3)) + \,{G_{4,{\phi}}} (- \,{G_{4}}^3 + 4 \,{G_{4,X}}^2 \,{X}^3 (\,{G_{4,X}} + 2 \,{G_{4,XX}} \,{X}) - 2 \,{G_{4}} \,{X}^2 (9 \,{G_{4,X}}^2 + 8 \,{G_{4,XX}}^2 \,{X}^2 + 4 \,{G_{4,X}} \,{X} (\,{G_{4,XX}} -  \,{G_{4,XXX}} \,{X})) + 2 \,{G_{4}}^2 \,{X} (3 \,{G_{4,X}} + \,{X} (9 \,{G_{4,XX}} + 2 \,{G_{4,XXX}} \,{X})))\right) \,\dot{\phi}\,.
\end{dmath}

\end{widetext}

\bibliographystyle{IEEEtran}
\bibliography{v4HorndeskiCCosmoPert}

\begin{thebibliography}{10}
\providecommand{\url}[1]{#1}
\csname url@samestyle\endcsname
\providecommand{\newblock}{\relax}
\providecommand{\bibinfo}[2]{#2}
\providecommand{\BIBentrySTDinterwordspacing}{\spaceskip=0pt\relax}
\providecommand{\BIBentryALTinterwordstretchfactor}{4}
\providecommand{\BIBentryALTinterwordspacing}{\spaceskip=\fontdimen2\font plus
\BIBentryALTinterwordstretchfactor\fontdimen3\font minus
  \fontdimen4\font\relax}
\providecommand{\BIBforeignlanguage}[2]{{%
\expandafter\ifx\csname l@#1\endcsname\relax
\typeout{** WARNING: IEEEtran.bst: No hyphenation pattern has been}%
\typeout{** loaded for the language `#1'. Using the pattern for}%
\typeout{** the default language instead.}%
\else
\language=\csname l@#1\endcsname
\fi
#2}}
\providecommand{\BIBdecl}{\relax}
\BIBdecl

\bibitem{nicolis2009galileon}
A.~Nicolis, R.~Rattazzi, and E.~Trincherini, ``Galileon as a local modification
  of gravity,'' \emph{Physical Review D}, vol.~79, no.~6, p. 064036, 2009.

\bibitem{Deffayet:2011gz}
C.~Deffayet, X.~Gao, D.~A. Steer, and G.~Zahariade, ``{From k-essence to
  generalised Galileons},'' \emph{Phys. Rev. D}, vol.~84, p. 064039, 2011.

\bibitem{horndeski1974second}
G.~W. Horndeski, ``Second-order scalar-tensor field equations in a
  four-dimensional space,'' \emph{International Journal of Theoretical
  Physics}, vol.~10, no.~6, pp. 363--384, 1974.

\bibitem{Kobayashi:2011nu}
T.~Kobayashi, M.~Yamaguchi, and J.~Yokoyama, ``{Generalized G-inflation:
  Inflation with the most general second-order field equations},'' \emph{Prog.
  Theor. Phys.}, vol. 126, pp. 511--529, 2011.

\bibitem{Rubakov:2016zah}
V.~A. Rubakov, ``{More about wormholes in generalized Galileon theories},''
  \emph{Theor. Math. Phys.}, vol. 188, no.~2, pp. 1253--1258, 2016.

\bibitem{Libanov:2016kfc}
M.~Libanov, S.~Mironov, and V.~Rubakov, ``{Generalized Galileons: instabilities
  of bouncing and Genesis cosmologies and modified Genesis},'' \emph{JCAP},
  vol.~08, p. 037, 2016.

\bibitem{Kobayashi:2016xpl}
T.~Kobayashi, ``{Generic instabilities of nonsingular cosmologies in Horndeski
  theory: A no-go theorem},'' \emph{Phys. Rev. D}, vol.~94, no.~4, p. 043511,
  2016.

\bibitem{Kolevatov:2016ppi}
R.~Kolevatov and S.~Mironov, ``{Cosmological bounces and Lorentzian wormholes
  in Galileon theories with an extra scalar field},'' \emph{Phys. Rev. D},
  vol.~94, no.~12, p. 123516, 2016.

\bibitem{Mironov:2019fop}
S.~Mironov, ``{Mathematical Formulation of the No-Go Theorem in Horndeski
  Theory},'' \emph{Universe}, vol.~5, no.~2, p.~52, 2019.

\bibitem{Akama:2017jsa}
S.~Akama and T.~Kobayashi, ``{Generalized multi-Galileons, covariantized new
  terms, and the no-go theorem for nonsingular cosmologies},'' \emph{Phys. Rev.
  D}, vol.~95, no.~6, p. 064011, 2017.

\bibitem{Creminelli:2016zwa}
P.~Creminelli, D.~Pirtskhalava, L.~Santoni, and E.~Trincherini, ``{Stability of
  Geodesically Complete Cosmologies},'' \emph{JCAP}, vol.~11, p. 047, 2016.

\bibitem{Mironov:2022quk}
S.~Mironov and A.~Shtennikova, ``{Stable cosmological solutions in Horndeski
  theory},'' \emph{JCAP}, vol.~06, p. 037, 2023.

\bibitem{Evslin:2011vh}
J.~Evslin and T.~Qiu, ``{Closed Timelike Curves in the Galileon Model},''
  \emph{JHEP}, vol.~11, p. 032, 2011.

\bibitem{Adams:2006sv}
A.~Adams, N.~Arkani-Hamed, S.~Dubovsky, A.~Nicolis, and R.~Rattazzi,
  ``{Causality, analyticity and an IR obstruction to UV completion},''
  \emph{JHEP}, vol.~10, p. 014, 2006.

\bibitem{Dubovsky:2005xd}
S.~Dubovsky, T.~Gregoire, A.~Nicolis, and R.~Rattazzi, ``{Null energy condition
  and superluminal propagation},'' \emph{JHEP}, vol.~03, p. 025, 2006.

\bibitem{Creminelli:2022onn}
P.~Creminelli, O.~Janssen, and L.~Senatore, ``{Positivity bounds on effective
  field theories with spontaneously broken Lorentz invariance},'' \emph{JHEP},
  vol.~09, p. 201, 2022.

\bibitem{Mironov:2023kzt}
S.~Mironov and M.~Valencia-Villegas, ``{Quartic Horndeski-Cartan theories in a
  FLRW universe},'' \emph{Phys. Rev. D}, vol. 108, no.~2, p. 024057, 2023.

\bibitem{Ageeva:2021yik}
Y.~Ageeva, P.~Petrov, and V.~Rubakov, ``{Nonsingular cosmological models with
  strong gravity in the past},'' \emph{Phys. Rev. D}, vol. 104, no.~6, p.
  063530, 2021.

\bibitem{Mironov:2018oec}
S.~Mironov, V.~Rubakov, and V.~Volkova, ``{Bounce beyond Horndeski with GR
  asymptotics and $\gamma$-crossing},'' \emph{JCAP}, vol.~10, p. 050, 2018.

\end{thebibliography}


\end{document}